# CHARACTERISATION OF AN ELECTROSTATIC VIBRATION HARVESTER

*T. Sterken[1,2], G. Altena[3], P.Fiorini[1], R. Puers[2]*

[1] IMEC, Leuven, Belgium, [2] K.U.Leuven, Leuven, Belgium
[3] IMEC – NL/Holst Centre, Eindhoven, The Netherlands
Tom.Sterken@imec.be

**ABSTRACT**

Harvesting energy from ambient vibration is proposed as an alternative to storage based power supplies for autonomous systems. The system presented converts the mechanical energy of a vibration into electrical energy by means of a variable capacitor, which is polarized by an electret. A lumped element model is used to study the generator and design a prototype. The device has been micromachined in silicon, based on a two-wafer process. The prototype was successfully tested, both using an external polarization source and an electret.

## 1. INTRODUCTION

Autonomy, mobility and lifetime are important characteristics of state-of-the-art electronic systems. The autonomy of these systems is assured by the use of primary or secondary batteries; the lifetime of storage based energy supplies depends on the ratio between the amount of energy stored in the battery to the power consumption of the electronic device. As the storage capacity is limited by the size (and weight) of the device, designers of electronic circuits make strong efforts to decrease the power consumption of the device.

An alternative approach is presented by ambient energy harvesters. These devices extract and convert energy from the environment. In this way the theoretical lifetime of autonomous systems is not limited by the size of the device, but depends on the availability of the ambient energy source. Low power design is still a key parameter in the autonomy of the system, as the power level of ambient energy supplies tends to vary in the lower microwatt ranges.

The feasibility of vibration powered systems depends on the ambient conditions, the envisioned application and boundary conditions such as maximum size, maximum weight and cost issues [1].

The generator presented in this contribution extracts energy from vibrations to which the autonomous device is subjected. The vibrations are characterised by small amplitudes of displacement in the micrometer scale and by relatively large frequencies, in the 100 - 3000 Hz range. These vibrations can be found in industrial environments. The generator design as presented is not suited for large amplitude motion scavenging, characterised by lower frequency vibrations. These are often related to human activities.

Vibration harvesting generators consist of a seismic mass whose motion is coupled either to an electromagnetic, a piezoelectric or an electrostatic transducer. The inertia of the mass acts as an artificial reference to the vibration. The mass is often suspended to reduce unwanted damping losses. At the same time the suspension can be designed to allow resonance at the main working frequency of the vibration. This approach is favourable for vibrations at higher frequencies with lower amplitudes: resonance converts the small amplitude of the package into an internal motion with a larger amplitude.

The work produced by the relative motion between the package of the device and the mass is converted to electrical energy by a transducer. In this work an electrostatic transducer is designed and fabricated using micromachining technology (Fig. 1). The transducer consists of a variable capacitor that is polarised. As the capacitance changes, a current is forced between the electrodes of the capacitance, through a load circuit. The current performs work in the electrical system, while the mechanical motion is damped by the associated electromechanical forces. The polarisation source can either be an external voltage source or a built-in electret.

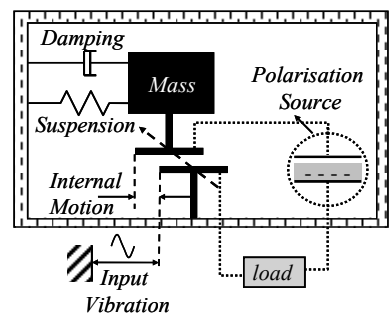

Figure 1. Schematic overview of an electrostatic harvester for vibrations.

## 2. DESIGN

The design of the prototype is based on the analysis of a lumped element model of the generator. This type of modeling converts the mechanical behavior to an electrical circuit, based on the equivalences between the behavior of masses, springs and dampers to inductances, capacitors and resistors respectively. The transducer then connects the mechanical equivalent circuit to the electrical load circuit. The equivalent model for the transducer is based on the relations between the forces, voltages, charges and geometry of the capacitance. As these relations are non-linear, a linearised small-signal model is used. The equivalent model consists of a transformer which is shunted by a capacitor (Fig. 2). The value of the capacitance corresponds to the value of the variable capacitor at equilibrium: $C_0$. The transformer is characterized by the ratio $n$ between the generated current $i$ and the velocity $\dot{z}$ at the mechanical poles of the generator:

$$n = \frac{i}{\dot{z}} = \frac{dC(z)}{dz} V_{pol} \qquad (1)$$

The analysis of this system leads to the following conclusions concerning design and optimization of a vibration harvester.

As a first rule the mass that is used as a reference should be maximized within the constraints of the application. A better reference will allow the use of a higher electromechanical damping. Furthermore the suspensions can be tuned to allow resonance at the dominant frequency of the vibration. At resonance the small input amplitude is converted to a larger amplitude, which is specifically useful at higher frequencies where the amplitude of the vibration is smaller than the system dimensions.

Although the use of resonance allows increasing the power that is generated, it also has a drawback: as the size of the system is minimized, resonance will result in impact of the mass onto the package. Besides reliability issues this impact also limits the output power of the system over a span of frequencies around the resonance frequency. The bandwidth depends on the ratio between the amplitude of the vibration and the maximum allowed displacement. The higher the maximum allowed displacement, the smaller the bandwidth will be. The optimum power will thus be reached when the mass and maximum displacement are tuned to the size constraints and energy spectrum of the application.

The design of the transducer allows tuning the output voltage and output impedance to the available vibration and mechanical design. In order to allow a maximum power transfer in a given maximum displacement, the *power factor* across the transducer should be optimized.

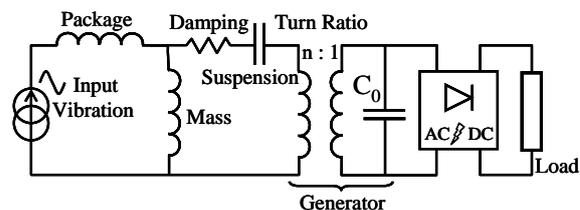

Figure 2. Equivalent electrical circuit

At the same time the electromechanical impedance should be as large as possible. These conditions are met when the load resistance equals the impedance of the capacitance. The electromechanical impedance is then given by:

$$Z_{EM} = \frac{n^2 R}{\sqrt{1+(\omega R C_0)^2}} = \frac{1}{\sqrt{2}} n^2 R = \frac{1}{\sqrt{2}} \frac{n^2}{\omega C_0} \qquad (2)$$

Optimal power is reached by maximizing equation 2. The design of the capacitance must allow a high capacitance change per micrometer of displacement.

The design consists of multiple variable overlap capacitors that are electrically connected in parallel. Scaling down the width and pitch of the electrodes will result in a higher transformation factor $n$ without influencing the total capacitance. This capacitance depends on the surface that is used (Figure 3c).

Note that the use of multiple capacitors in parallel increases the frequency of the generated electrical signal: once the motion of the mass exceeds the pitch of the capacitors, the electrical signal is reproduced. This effect partly counteracts the advantage on $n$ versus $C_0$ of the structure: at higher electrical frequency the capacitance will shunt the load circuit.

## 3. FABRICATION

The variable capacitor is fabricated using a stack of two bonded wafers. The fixed electrode is created on a Pyrex wafer by sputtering and patterning an aluminum layer. Then photosensitive BCB (Benzo-Cyclo-Butene) by the Dow Chemical Company is spun and patterned (Figure 3a). Both the thickness of the aluminum layer (200 nm) and the BCB layer (1.2 µm) determine the distance $d$ between the fixed electrode and the moving electrode. The use of Pyrex allows reducing parasitic capacitances shunting the generator, while the thermal expansion coefficient remains tuned to the one of silicon. This requirement is a necessity as BCB bonding occurs at 250°C.

The moveable electrode of the capacitor is bulk micromachined together with the mass and the suspension springs in a 6" silicon <100>-wafer. First, the capacitor, the mass and the springs are patterned and then etched into the silicon by deep reactive ion etching (DRIE) using a STS ASE etcher to a depth of 100 µm. Next, a 500 nm thick $SiO_2$ layer is grown using thermal oxidation.
A 150 nm thick $Si_3N_4$ layer is used on the back side of the wafer as a hard mask for wet anisotropic etching (Figure 3b). The two wafers are then combined using aligned wafer-to-wafer bonding where the BCB layer acts as an adhesive [3]. This two-wafer stack is anisotropically etched in KOH. A layer of approximately 30 µm is left to prevent leakage of KOH into the cavity between the Si wafer and the glass wafer (Figure 3c). This 30 µm thick Si layer is removed with a DRIE process. The variable capacitor is finally released by removing the $SiO_2$ which served as an etch stop. SEM micrographs of the mass and suspension are shown in Figure 4.

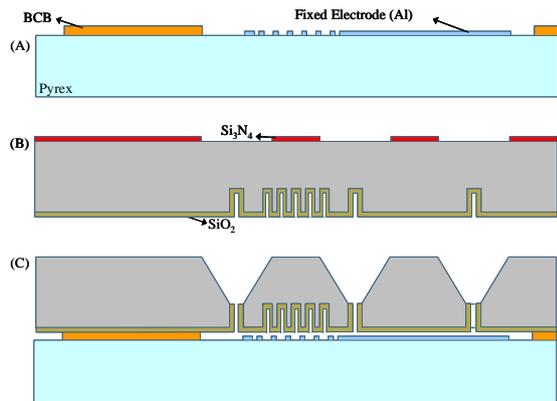

Figure 3. The variable capacitor consists of two wafers (A+B), which are next bonded and released (C).

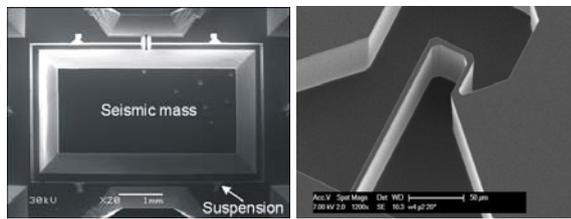

Figure 4. SEM photographs of the released device (left) and a close-up of the suspension (right).

## 3. CHARACTERISATION

During a first test the variable capacitance was measured as a function of displacement (Figure 5). A small discrepancy can be noticed in the capacitance change per micrometer. From these curves the capacitance change per micrometer is calculated and given in figure 5. The second test in the characterisation of the prototype consists in displacing the mass over a 40 µm displacement at a frequency of 40 Hz.

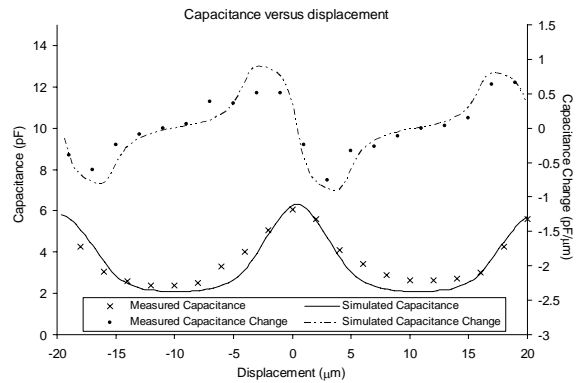

Figure 5. Capacitance and capacitance change per unit of displacement, as measured and simulated.

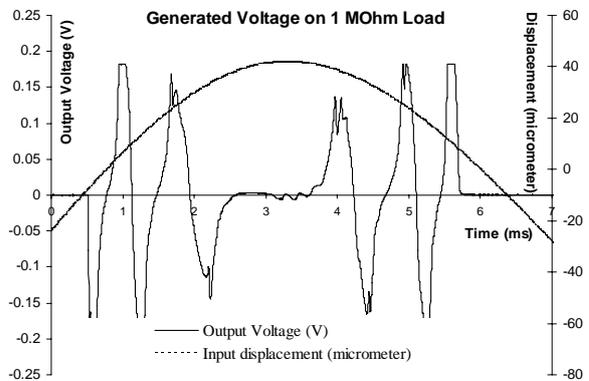

Figure 6. Voltage generated as a function of displacement over a 1 MΩ load resistor.

In this way the mass moves across multiple capacitances per cycle. The shape of the signal corresponds to the simulated and measured changes of capacitance versus displacement. During this test the polarisation voltage was limited to 10V, as higher voltages resulted in a pull-in of the mass. In figure 6 the generated signal is illustrated as a function of time, as well as the displacement of the probe. The signal is symmetric towards the maximum displacement point. Peak voltages up to 200mV are generated over a 1 MΩ load, while the velocity of the mass has a maximum of 0.03 m/s.
A final test applied to the device consists in charging the $SiO_2$ layer between the two electrodes of the variable capacitor. The electret is charged by electrostatically pulling down the movable electrode until the electret layer touches the aluminum electrode. The electret is charged up to the pull-in voltage, which was 15V. Unfortunately this test does not allow checking the electret voltage after charging without destroying the device.

As the mass of the generator was displaced by the probe, a similar voltage pattern was generated as given in Figure 6.

The device was next mounted on an industrial tool with a strong vibration of 1g at 2.6 kHz (Figure 7). As the displacement of the mass does not exceed the pitch of the electrodes, the signal is sinusoidal at the frequency of the vibration. Due to a small misalignment of the two electrodes, the shape of the generated voltage on the positive half of the cycle differs from the shape on the negative half of the cycle. A power of 90 nW is generated on a 1 MΩ load resistor.

### 4. CONCLUSIONS

An electrostatic generator based on ambient vibrations was studied and designed according to the conclusions of the analysis. A process flow is developed based on a stack of two bonded wafers, each bearing an electrode of the variable capacitor. After fabrication the generator is tested, first by displacing the mass, next by subjecting the device to industrial vibrations. The capacitor proved to extract energy, both when polarised by an external voltage source as well as by means of an electret.

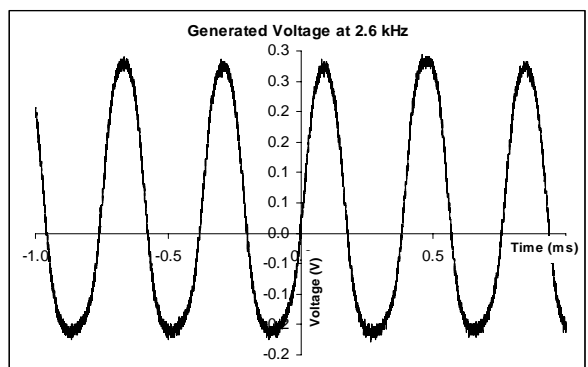

Figure 7. A voltage of 300mV is generated across a 1 MΩ resistor when subjected to a 2.6kHz vibration at 1g.